\documentclass{aipproc}
\layoutstyle{6x9}

\begin{document}
\title{New Physics and novel Higgs signals}
\author{ J.~Lorenzo~Diaz-Cruz}{address={Instituto de Fisica, BUAP, 72570 Puebla, Pue, Mexico
}}

\begin{abstract}
We review some of the results of our recent work dealing with
the novel type of Higgs signals that arise when one
considers extensions of the standard model.
 We discuss first possible deviations on
the Higgs couplings due to heavy particles, in the context of the
MSSM and with large extra-dimensions. Then, we present several
models where it is possible to induce flavor violating Higgs
couplings, and probe them at future hadron colliders through the
LFV Higgs decay $h\to \tau\mu$ or with rare top decays.
\end{abstract}

\maketitle

\section{Introduction}

The discovery of the Higgs boson is certainly one of the most
cherished goals of present and future high-energy experiments. In
fact, the reported bounds on the Higgs boson mass \cite{hsmradc},
together with data on neutrinos \cite{superkam} and CP-violation
with B-mesons \cite{nirtalk}, can be considered some of the most
important recent results in Particle Physics, and are already
helping us to shape our understanding of flavor physics and
electroweak (EW) symmetry breaking. The Higgs boson mass is
constrained by radiative corrections to lay in the range 110-185
GeV at 95 \% c.l. \cite{hsmradc}; such a light Higgs boson ($h$)is
consistent with the  prediction of weak scale SUSY, which has
become on of the preferred extensions of the SM \cite{susyrev}.

 The characteristic Higgs boson couplings determine the
strategies employed for its search at present and future
colliders \cite{hixhunter}.
For instance, the Higgs-fermion couplings can be
studied by open production of $t\bar{t}h, b\bar{b}h$ at
hadron colliders or NLC. However, the possible presence of heavy
particles associated with physics beyond the SM, can induce corrections
to such couplings, which can modify the SM
predictions for Higgs production or decays.
Heavy particles can also induce tree-level corrections, as it occurs
in scenarios with large extra-dimensions, where the KK modes can
contribute to the associated production of Higgs with Z boson
at NLC, as it will be discussed next.

Flavor violation is another phenomena that could be
tested in the Higgs sector.  The most widely studied scenarios for Higgs
searches, assume that the Flavor-Conserving (FC) Higgs-fermion couplings
only depend on the diagonalized fermion mass matrices, while flavor-violating
(FV) Higgs transitions are absent or highly suppressed \cite{Krawczyk:1998ru}.
 Indeed, within the SM the Higgs boson-fermion couplings are only sensitive
to the fermion mass eigenvalues.  However, if one considers extensions
of the SM, it is possible to induce new flavored Higgs interactions.
These new interactions could be tested through the
lepton-flavour violating (LFV) Higgs decays, such as
$h\to \tau\mu/ \tau e$ \cite{myhixLFV},
which can reach detectable levels in several models that
will be discussed next;  similarly, these FV scalar interactions can also
be tested with the rare decays of the top quark.

\section{Deviations of the Higgs couplings and heavy
particles}

Within the minimal standard model (SM), a light Higgs boson mass
is favored by present data \cite{hsmradc}. However one can show
that the effect of new physics can modify the Higgs couplings with
gauge bosons, in a manner that cancels the (virtual) Higgs
contribution that appears in the analysis of EW radiative
corrections, and thus allow heavier Higgs masses, up to about 600
GeV, that still could be accessible at future colliders. This was
first proved using a model-independent effective lagrangian
approach in Refs. \cite{Meandthem}. Nevertheless, even after a
Higgs signal will be seen, probably at the Tevatron and/or LHC, it
will become crucial to measure its mass, spin and couplings, to
elucidate its nature. In particular, the Higgs coupling to light
fermions ($b\bar{b}, c\bar{c}, \tau^+\tau^-$) could be measured at
next-linear collider (NLC) with a precision of a few percent
\cite{hixcoupl}, which can be used to constrain physics beyond the
SM.
 For instance, higher-dimensional operators of the type
$\Phi^\dagger \Phi \bar{Q}_L \Phi b_R$
involving the third family,  will generate corrections
to the coupling $h\bar{b}b$, which in turn will
modify the dominant decay of the light Higgs,
as well as the associated production of the Higgs with
b-quark pairs. This effect was studied within the minimal SUSY SM (MSSM),
and found to be  detectable in the large $\tan\beta$ regime
\cite{ourhixbb,carena}. The Higgs sector of the MSSM
includes two Higgs doublets, and the light Higgs boson
(with mass bound $m_h \leq 125$ GeV), is perhaps the strongest
prediction of the model.

Any additional heavy particle that receives its mass from
the SM Higgs mechanism, will also contribute to the
1-loop vertices, an effect that can be tested
through Higgs production , such as the
loop-induced gluon fusion production of Higgs bosons at
hadron colliders, or through the decay into photon pairs
($h\to \gamma\gamma$). However, such heavy particles will
also induce non-decoupling corrections to the tree vertices
$hf\bar{f}$, $hWW$ and $hZZ$, which can
affect the decay rate of detectable signatures \cite{myhixm}.

Heavy particles can also induce tree-level corrections, as it occurs
in scenarios with large extra-dimensions, which were
proposed as an alternative solution to the hierarchy
problem \cite{newexd}. The location of the Higgs  can be studied within
the context of a model with two-Higgs doublets, one living in the brane,
while the other penetrates the bulk. Due to this assignment, new vertices
of the type $hZZ_{KK}$ between the higgs (h), the Z boson and its KK
resonances ($Z_{KK}$), arise in the model. These vertices in turn affect
the Higgs phenomenology. For instance we found that the KK states
contribute to the associated production of Higgs with Z boson, in a manner
that suppresses the Higgs rates at LEP and Tevatron, while giving large
enhancements on the cross-section at the future
LHC and NLC \cite{ourhixdm}.

\section{Flavor-violating Higgs couplings}

New flavored Higgs interactions can be induced when one considers
extensions of the SM, which either present a significant new
source of flavor-changing  transition  or are aimed precisely to
explain the pattern of masses and mixing angles of the quarks and
leptons. Namely, when  additional fields that have non-aligned
couplings to the SM fermions, i.e. which are not diagonalized by
the same rotations that diagonalize the fermion mass matrices, and
also couple to the Higgs boson, then such fields could be
responsible for transmitting the structure of the flavor sector to
the  Higgs bosons, thereby producing {\it{a more flavored Higgs
boson}} \cite{myMFHB}. Depending on the nature of such new
physics, we can identify two possibilities for flavor-Higgs
mediation, namely:

\begin{itemize}
\item {\it{RADIATIVE MEDIATION.}} In this case the Higgs sector has
diagonal couplings to the fermions at tree-level.
However, the presence of new particles associated with extended
flavor physics, which couple both to the Higgs and to the SM fermions,
will induce corrections to the Yukawa
couplings and/or new FCNC process at loop levels.
Within the MSSM, it can be shown that flavor-Higgs mediation
is of radiative type, and it communicates the non-trivial flavor
structure of the soft-breaking sector  to the  Higgs bosons
through gaugino-sfermion loops.

\begin{enumerate}
\item As an illustration of the SUSY case, we have evaluated the
slepton-gaugino contributions to the LFV Higgs-lepton vertices,
with slepton mixing originating from the trilinear $A_l$-terms.
The slepton mixing is constrained by the low-energy data, but it
mainly suppress the FV's associated with the first two
family sleptons, and still allows the flavor-mixings
between the second- and third-family sleptons,
to be as large as $O(1)$. Thus, the general $6\times 6$
slepton-mass-matrix can be reduced down to a $4\times 4$ matrix,
involving only the smuon and stau sectors.
Such pattern of large slepton mixing, and the resulting $h\tau\mu$
coupling can also be motivated by the large neutrino
mixing observed with atmospheric neutrinos \cite{superkam}.
For this pattern of large trilinear A-terms, we find that
bounds on $\tau-\mu$ transitions, allow the decay mode $h\to \tau \mu$,
to reach a B.R. of the order $10^{-4}$, which enters into the domain
of detectable signals \cite{prihmutau}.

\item This reduction of sfermion mass matrices was discussed first for
the squarks in ref. \cite{oursqmix}.  Similar mixing between the
stop and scharm squarks, can be tested through the rare decays
of the top quark \cite{mytcgama}; for instance one finds that
the decay  $t\to c h$ reaches branching ratios of the order
$10^{-3}-10^{-4}$, which could be detected
at future hadron colliders, such as LHC.

\end{enumerate}

\item {\it{MIXING MEDIATION.}} Modifications to the Higgs-flavor
structure can also arise when additional particles (bosons or fermions)
mix with the SM ones. These new interactions could then be transmitted
to the Higgs sector, either through scalar-Higgs mixing or through
mixing of SM fermions with exotic ones. Some models where
this flavor-violating Higgs interactions appear are:

\begin{enumerate}
\item In the general Two-Higgs doublet model (THDM-III)
\cite{hifcnc,mythdiiia}, where large FV Higgs couplings are allowed,
we obtain $B.R.(H\to \tau \mu / \tau e) \simeq 10^{-1}$
\cite{myhixLFV},
which can be detected at Tevatron Run-II and LHC
\cite{prihmutau}.

\item Along these lines we have also considered a multi-Higgs $E_6-$inpired
model, supplemented with an abelian flavor symmetry, where similar
rates for $h \to \tau\mu$ are obtained \cite{myMFHB}.

\item We have also consider mixing between SM and exotic fermions,
within the context of a LR model with mirror fermions \cite{hlfvLRMM}
as a source of LFV , when this is transmitted to the Higgs sector, it
induces the decay $h\to \tau \mu$ at detectable rates.

\end{enumerate}

\end{itemize}

Thus, as a consequence of the presence of LFV Higgs interactions,
the decay $h \to \tau\mu$ can be induced at rates that could be
detected at future colliders, such as Tevatron and LHC. Large LFV
Higgs couplings could be the manifestation of a deeper link
between the Higgs and flavor sectors.

\section*{Acknowledgements}

This work was presented at the X Mexican School of particles and
fields, dedicated to my professors and friends, Augusto Garcia and
Arnulfo Zepeda, pioneers of the field in Mexico and Latinamerica.
Work supported by CONACYT-SNI (Mexico).

\end{document}